\documentclass[12pt,preprint]{aastex}

\usepackage{amsmath}
\renewcommand{\thefootnote}{\fnsymbol{footnote}}

\begin{document}

\title{Water Absorption from Gas Very Near the Massive Protostar AFGL 2136 IRS 1\footnote{Based on observations collected at the European Organisation for Astronomical Research in the Southern Hemisphere, Chile as part of program 089.C-0321}}

\author{Nick Indriolo\altaffilmark{1},
D.~A.~Neufeld\altaffilmark{1},
A.~Seifahrt\altaffilmark{2},
M.~J.~Richter\altaffilmark{3}
}

\altaffiltext{1}{Department of Physics and Astronomy, Johns Hopkins University, Baltimore, MD 21218}
\altaffiltext{2}{Department of Astronomy and Astrophysics, University of Chicago, Chicago, IL 60637}
\altaffiltext{3}{Department of Physics, University of California Davis, Davis, CA 95616}

\begin{abstract}

We present ground-based observations of the $\nu_1$ and $\nu_3$ fundamental bands of H$_2$O toward the massive protostar AFGL~2136~IRS~1, identifying absorption features due to 47 different ro-vibrational transitions between 2.468~$\mu$m and 2.561~$\mu$m.  Analysis of these features indicates the absorption arises in warm ($T=506\pm25$~K), very dense ($n({\rm H}_2)>5\times10^{9}$~cm$^{-3}$) gas, suggesting an origin close to the central protostar.  The total column density of warm water is estimated to be $N({\rm H_{2}O})=(1.02\pm0.02)\times10^{19}$~cm$^{-2}$, giving a relative abundance of $N({\rm H_{2}O})/N({\rm H}_2)\approx10^{-4}$.  Our study represents the first extensive use of water vapor absorption lines in the near infrared, and demonstrates the utility of such observations in deriving physical parameters.

\end{abstract}


\section{INTRODUCTION} \label{section_intro}

\setcounter{footnote}{3}
\renewcommand{\thefootnote}{\arabic{footnote}}

Water is expected to be one of the most abundant species in warm, dense regions of the interstellar medium where its formation is facilitated by neutral-neutral reactions.  In the innermost regions of protostellar envelopes, chemical models predict that nearly half of the total oxygen abundance is in the form of gas-phase H$_2$O, $n({\rm H_2O})/n({\rm H_2})\approx10^{-4}$, while the other half is in CO \citep{ceccarelli1996}.
Because the relative populations in excited rotational states of H$_2$O are highly dependent on physical parameters (e.g., temperature, density, radiation field), water can be a powerful diagnostic of the conditions in protostellar environments.    

Due to abundant water in Earth's atmosphere, ground-based observations of astrophysical water vapor have been rather limited in scope, primarily focusing on maser emission.  Indeed, the first detection of interstellar water was made by observing maser emission from the $6_{1,6}$--$5_{2,3}$ transition of H$_2$O near 22~GHz \citep{cheung1969}.  Since then, ground-based observations of water masers in star forming regions have expanded to include the following transitions: $10_{2,9}$--$9_{3,6}$ at 321~GHz \citep{menten1990_321GHz}; $5_{1,5}$--$4_{2,2}$ at 325~GHz \citep{menten1990_325GHz}; $3_{1,3}$--$2_{2,0}$ at 183~GHz \citep{cernicharo1990}; $6_{4,3}$--$5_{5,0}$ at 439~GHz \citep{melnick1993}; $6_{4,2}$--$5_{5,1}$ at 471~GHz \citep{melnick1993}.  Of the various water maser transitions, that at 22~GHz is by far the most frequently observed, with detections in over 1000 different sources \citep{benson1996,valdettaro2001,sunada2007}.  These H$_2$O masers are good tracers of high-mass star formation, likely arising in dense, shocked gas associated with molecular outflows.  

Ground-based observations of water in non-maser transitions are extremely limited, with only a few studies detecting emission in protoplanetary disks from transitions in the near-IR \citep{najita2000,carr2004,salyk2008} and mid-IR \citep{pontoppidan2010}.  Instead, observations of astrophysical water in non-maser transitions have relied mostly on space-based observatories.  ISO-SWS (Infrared Space Observatory-Short Wavelength Spectrometer) revealed water absorption near 6~$\mu$m from the $\nu_2$ ro-vibrational band toward massive protostars \citep{boonman2003}.  {\it Spitzer}-IRS (InfraRed Spectrometer) observations showed H$_2$O emission in pure rotational transitions across the mid-infrared in protoplanetary disks \citep[e.g.,][]{pontoppidan2010spitzer}.  Most recently, {\it Herschel}-HIFI (Heterodyne Instrument for the Far-Infrared) has enabled observations of water in several of the lowest rotational states, including the ground state \citep[e.g.,][]{vandishoeck2011,sonnentrucker2010}.  Unfortunately, the current lack of space-based observatories capable of observing water threatens to stagnate our investigations of astrophysical H$_2$O unless new methods are devised.  Here, we present the first extensive study of the $v=1$--0, $\nu_1$ and $\nu_3$ bands of water---detected in absorption toward AFGL~2136~IRS~1 via ground-based observations in the near-IR while targeting the $v=1$--0, $R(0)$ transition of HF at 2.499385~$\mu$m \citep{indriolo2013_HF}---and examine the potential of such observations for future studies.

\section{TARGET CHARACTERISTICS} \label{section_source}

AFGL~2136 is a star-forming region at a distance of about 2~kpc that shows a complex morphology at near infrared ($H$-band and $K$-band) wavelengths due to dust grains reflecting light from the central source, IRS~1, an embedded massive protostar with luminosity $L\sim5\times10^4$~$L_{\sun}$ \citep{minchin1991,kastner1992,murakawa2008}.   Polarization measurements are indicative of a circumstellar disk at a position angle of about 45$^\circ$ \citep{kastner1992,murakawa2008}, and emission maps in CO show massive molecular outflows oriented perpendicular to the proposed disk \citep{kastner1994}.  Gas velocities in the outflow lobes 
suggest a modest---less than 40$^\circ$---inclination with respect to the plane of the sky.  A hypothetical picture of AFGL~2136~IRS~1 based on mid-infrared interferometry shows an envelope, torus, and disk component \citep{dewit2011}.

Infrared spectra reveal both warm and cold gas along this line of sight.  Absorption due to gas-phase CO \citep{mitchell1990co} and H$_2$O \citep{boonman2003} in highly excited states is observed, along with ice bands of H$_2$O, CO$_2$, and CH$_3$OH \citep{willner1982,vandishoeck1996CO2,brooke1999,gerakines1999}.  The warm gas is thought to be in the core surrounding the massive protostar, while the cold gas is in the outer envelope or molecular cloud from which the protostar collapsed.  Observations of CO, CS, and H$_2$CO in emission give a systemic velocity of $v_{\rm LSR}=22.8$~km~s$^{-1}$ for the molecular cloud \citep{vandertak2000yso}.  A small (3--5~km~s$^{-1}$) redshift with respect to the systemic velocity distinguishes the warm gas from the cold cloud.

\section{OBSERVATIONS \& DATA REDUCTION} \label{section_obs_rx}

\begin{figure}
\epsscale{1.0}
\plotone{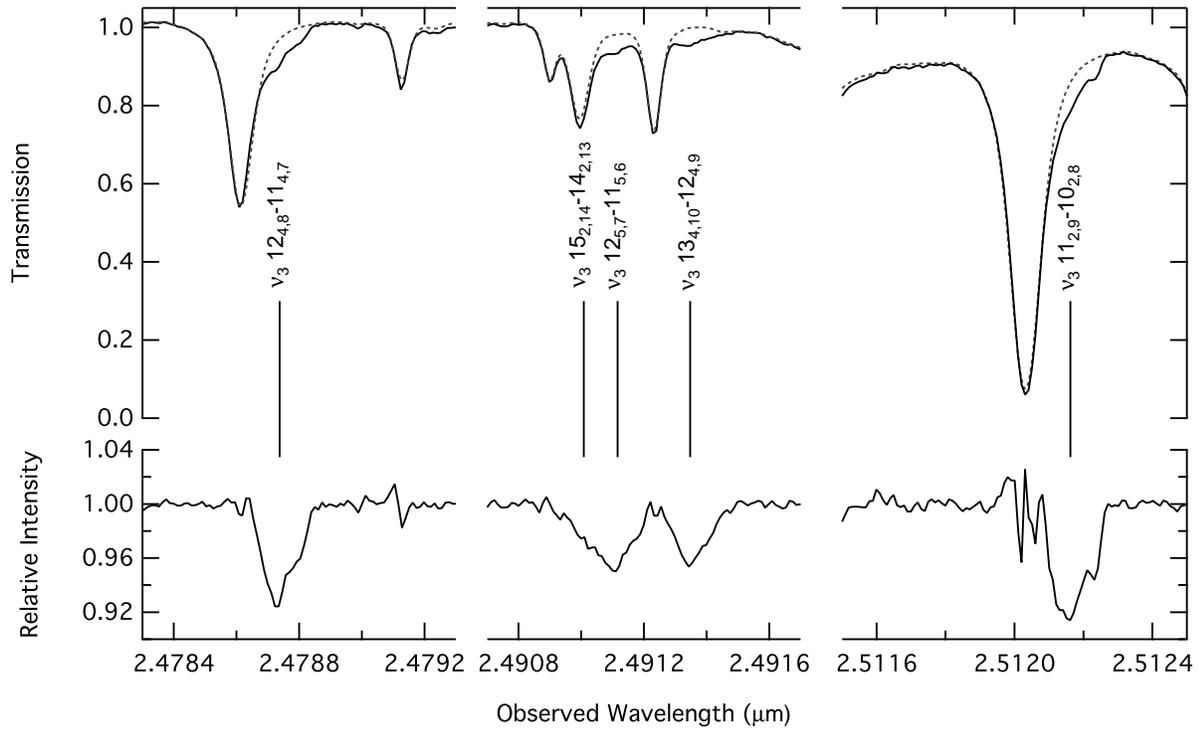}
\caption{Select wavelength regions showing the observed spectra (solid) and model atmospheric spectra (dashed) above, and the observed divided by model spectra below.  In all cases there is absorption at about +15~km~s$^{-1}$ (marked by vertical lines) with respect to the atmospheric water features.}
\label{fig_absorption}
\end{figure}

AFGL~2136~IRS~1 was observed on July 6, 2012 for a total of 3960~s using the Cryogenic High-resolution Infrared Echelle Spectrograph \citep[CRIRES;][]{kaufl2004} on UT1 at the Very Large Telescope.  Observations were performed in service mode, and CRIRES was used with its 0\farcs2 slit to provide a resolving power (resolution) of about 100,000 (3~km~s$^{-1}$).  A reference wavelength of 2502.8~nm set the wavelength ranges on detectors 1 through 4 to be 2.4679--2.4796~$\mu$m, 2.4831--2.4943~$\mu$m, 2.4974--2.5081~$\mu$m, and 2.5111--2.5212~$\mu$m, respectively.  Due to the lack of a bright natural guide star, the adaptive optics system was not utilized.  The slit was oriented at a position angle of 45$^\circ$ (i.e., along a northeast-to-southwest axis) to minimize interference from surrounding nebulosity.  Spectra were obtained in an ABBA pattern with 10\arcsec\ between the two nod positions and $\pm$3\arcsec\ jitter width.  Details regarding the data reduction procedure can be found in \citet{indriolo2013_HF}.  Processed one-dimensional science spectra were divided by model atmospheric spectra \citep{seifahrt2010} for the purpose of removing atmospheric absorption lines.  Science spectra, model atmospheric spectra, and the resulting ratioed spectra in select wavelength regions are shown in Figure \ref{fig_absorption}.

\section{RESULTS \& ANALYSIS} \label{section_results}

\begin{figure}
\epsscale{1.0}
\plotone{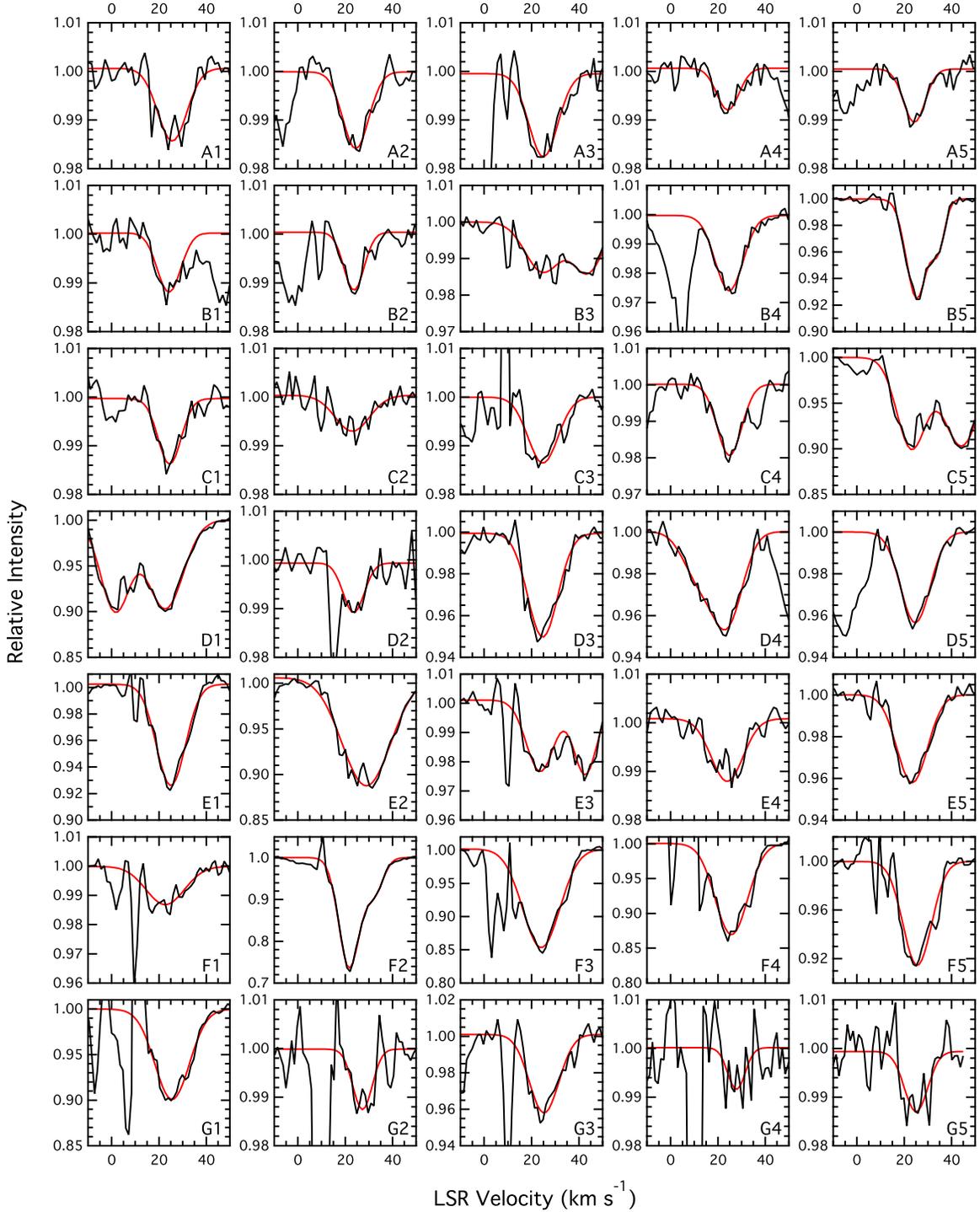}
\caption{Spectra showing all 35 absorption features found in our analysis.  Transition identifications for each panel are given in Table 1.  Note the different relative intensity scales for each panel.  In many cases, features near 10~km~s$^{-1}$ are due to incomplete removal of telluric water lines.}
\label{fig_spectra}
\end{figure}

\begin{figure}
\epsscale{0.8}
\plotone{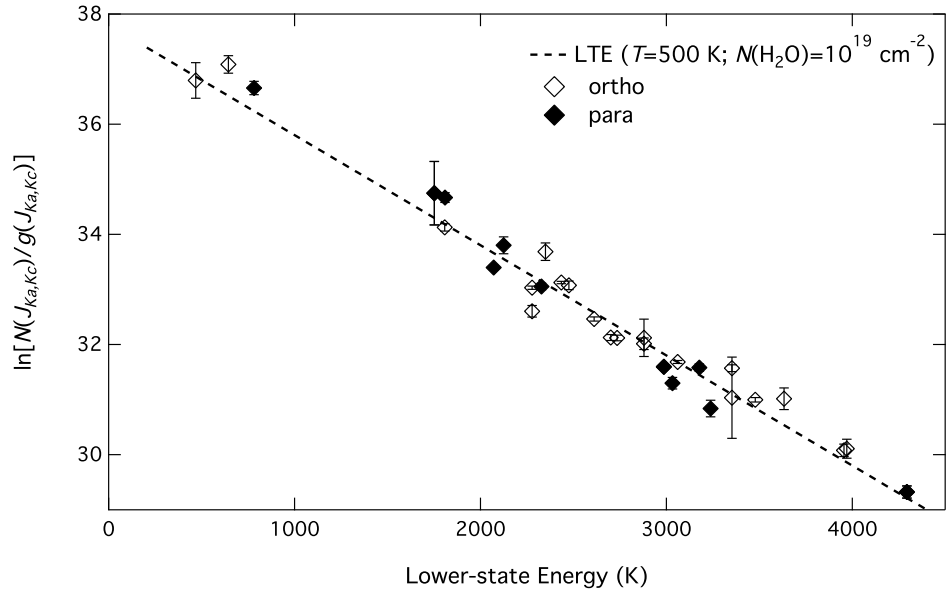}
\caption{Rotation diagram for unblended transitions where column densities could be determined.  Ortho levels are given by open diamonds, and para levels by filled diamonds.  The dashed line is {\it not} a fit to the data, but marks the expected populations for LTE at $T=500$~K with $N({\rm H_2O)}=10^{19}$~cm$^{-2}$.}
\label{fig_rotdiag}
\end{figure}

In the full wavelength range covered by our observations we identified 35 separate absorption features arising from 47 different transitions of the $v=1$--0 $\nu_1$ (symmetric stretch), $v=1$--0 $\nu_3$ (asymmetric stretch), and $v=2$--0 $\nu_2$ (bend) bands of H$_2$O.  All of these are shown in Figure \ref{fig_spectra} in velocity space, along with Gaussian fits to the absorption features.  Fits are comprised of either one or two Gaussian components, with two components used in cases where there are multiple features (e.g., panels D1 \& E3), or the main feature has a prominant ``shoulder'' on the right-hand side (e.g., panels B5 \& F2).  Parameters extracted from the fitting procedure---LSR velocity, velocity FWHM, equivalent width---are given in Table \ref{tbl_absorption}, along with transition labels, column densities, and a key denoting which transitions are shown in each panel in Figure \ref{fig_spectra}.  The mean LSR velocity is $24.6\pm1.1$~km~s$^{-1}$, and the mean velocity FWHM is $13.6\pm2.5$~km~s$^{-1}$.  A rotation diagram (Figure \ref{fig_rotdiag}) was made using column densities reported in Table \ref{tbl_absorption}.  The relatively straight line traced out by the data in $\ln(N/g)$ versus lower state energy (where $g$ is the statistical weight) is indicative of gas with a single temperature in local thermodynamic equilibrium (LTE).

To determine the physical conditions in the absorbing gas, we have modeled the excitation of the lowest 120 rotational states of ortho- and para-water using a statistical equilibrium code to determine the expected level populations.  In this analysis, we used an escape probability method to treat the effects of radiative trapping.  For the collisional excitation of H$_2$O by H$_2$, we adopted the rate coefficients computed recently by \citet{daniel2011}, and then used an artificial neural network method \citep{neufeld2010_nn} to extrapolate these to states of higher energy than those for which calculations were available.   We also included the effects of radiative pumping---in both pure rotational and ro-vibrational transitions---by the infrared continuum radiation emitted by AFGL~2136~IRS~1; here, we adopted the spectral energy distribution for AFGL~2136~IRS~1 implied by a fit to available data obtained by \citet[][the TO model plotted in their Figure 7]{murakawa2008}.

\begin{figure}
\epsscale{0.8}
\plotone{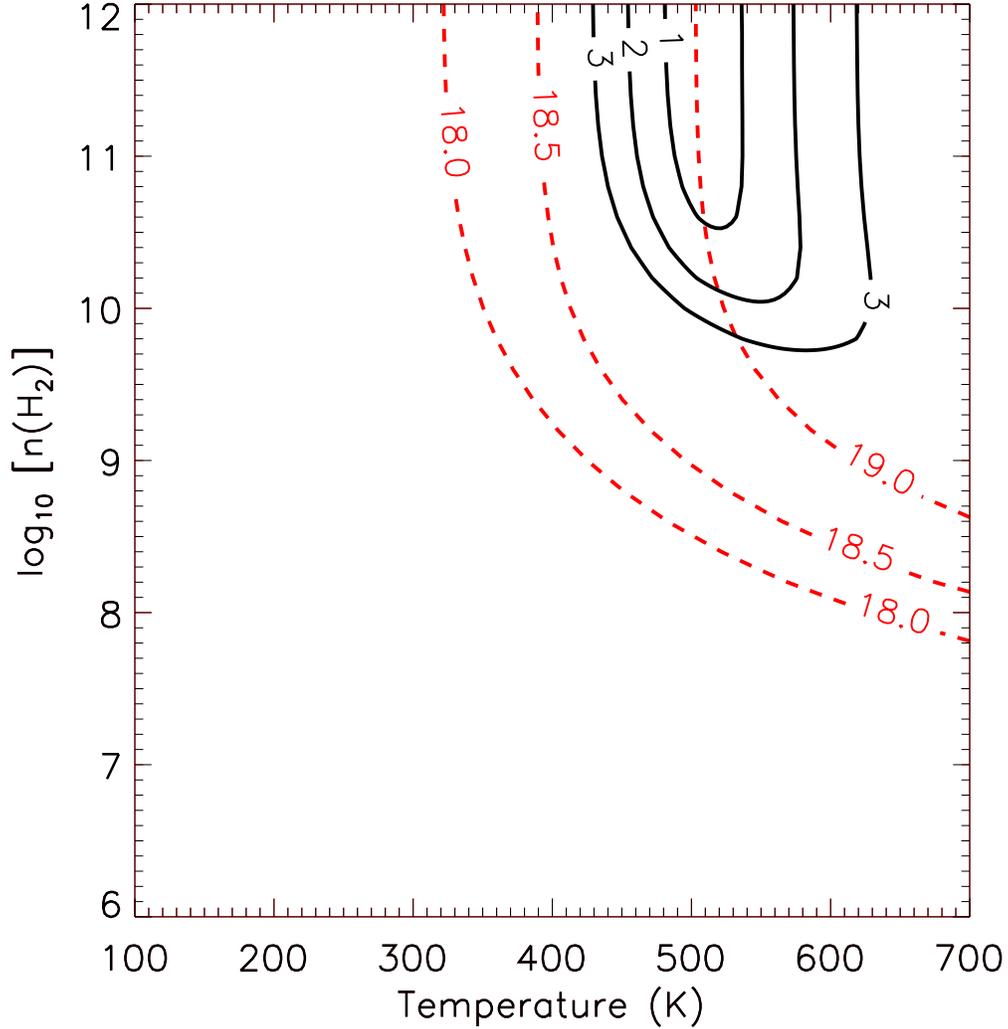}
\caption{
Best fit temperature, density, and H$_2$O column density as determined from a statistical equilibrium analysis of the level populations given in Table 1.   Solid black contours show the quantity $(\chi^2 - N_{\rm dof})^{1/2}$, where $N_{\rm dof}$ is the number of degrees of freedom (i.e., the number of independent data points minus the number of fitted parameters).  These contours represent 68.3\%, 95.4\% and 99.7\% confidence limits.   Here, the $\chi^2$ has been scaled (downward) to make its minimum value equal $N_{\rm dof}$.  The required reduction factor suggests that the statistical error estimates in Table \ref{tbl_absorption} are smaller than the actual typical errors by a factor 4.29, a discrepancy that presumably reflects the presence of systematic errors.  Dashed red contours show the values of ${\rm log}_{10} [N({\rm H_2O})/{\rm cm}^{-2}]$ needed to fit the data.  The statistical equilibrium analysis presented here includes the effects of radiative trapping of far-infrared water transitions for an assumed value of $10^{18}$~ cm$^{-2}$ per km~s$^{-1}$ for $N({\rm H_2O})/\Delta v$.}
\label{fig_chi2}
\end{figure}

The importance of radiative excitation depends upon the assumed distance, $d$, of the absorbing material from the source of continuum radiation.  For $d > 200$~AU, radiative excitation has a negligible effect upon the population of the rotational states that we have observed; in this limit, the observed level populations place strong constraints upon the temperature, H$_2$ density, and total water column density.  A chi-squared minimization yields a best-fit gas temperature of $506\pm25$~K, a best-fit H$_2$O column density of $(1.02\pm0.02)\times10^{19}$~cm$^{-2}$, and places a $3\sigma$ lower limit of $5\times10^{9}$~cm$^{-3}$ on the H$_2$ density.\footnote{Note that $N({\rm H_2O})$ is dependent on the single temperature model utilized and is likely not as tightly constrained as the quoted uncertainties suggest.}  In Figure \ref{fig_chi2}, we show the contours of chi-squared in the temperature-density plane, obtained in the limit $d > 200$~AU.  For $d < 200$~AU, the effects of radiative pumping reduce our lower limit on the H$_2$ density by up to an order of magnitude, the weakest constraints being obtained for $d\sim120$~AU.

Taking the range of warm H$_2$ column densities previously determined \citep[$N({\rm H}_2)=(5.5$--$9.8)\times10^{22}$~cm$^{-2}$;][]{indriolo2013_HF}, we find a relative abundance of $N({\rm H_2O})/N({\rm H_2})=(1.0$--$1.9)\times10^{-4}$, in relatively good agreement with predictions made by chemical models for warm, dense gas near massive protostars \citep[e.g.,][]{ceccarelli1996,doty2002}.  The lower limit on the number density can be used with the total H$_2$ column density to estimate the size of the region where the H$_2$O absorption occurs.  Adopting the larger value of $N({\rm H}_2)=9.8\times10^{22}$~cm$^{-2}$, the density limit of $n({\rm H}_2)>5\times10^{9}$~cm$^{-3}$ results in a path-length limit of $L<2\times10^{13}$~cm (1.3~AU).  The weaker density constraints in the scenario where radiative pumping is important relax this size limit to $L\lesssim13$~AU.  Given the high density, high temperature, and small path length, this gas must be located fairly close to the central protostar, likely in the inner envelope or a circumstellar disk.

\section{DISCUSSION}

\subsection{Comparison with Observational Results}

The physical parameters derived from our analysis can be compared to those found by previous studies of AFGL~2136~IRS~1.  Water absorption has been observed in this source before via the $v=1$--0 $\nu_2$ band near 6~$\mu$m using ISO-SWS \citep{vandishoeck1996,boonman2003}.  The lower spectral resolution of that instrument ($\lambda/\Delta\lambda\sim1400$) made it so that all absorption features were blends of several different transitions.  In order to determine best-fit parameters, \citet{boonman2003} produced a suite of synthetic spectra arising from different combinations of $T$, $N({\rm H_2O})$, and $b$ (Doppler line width), and compared them to the observed spectrum.  Adopting a value of $b=5$~km~s$^{-1}$, they found $T=500^{+250}_{-150}$ and $N({\rm H_2O})=(1.5\pm0.6)\times10^{18}$~cm$^{-2}$.  The excitation temperature is in excellent agreement with our findings, but their water column density is about 7 times smaller.  This result was obtained with a model only considering absorption, and \citet{boonman2003} postulated that emission filling in some of the absorption features could result in column densities being underestimated by factors of 3--6.

It is unclear to what extent emission competes with absorption in the H$_2$O spectra presented by \citet{boonman2003}.  If a spherical cloud lies entirely within the telescope beam (slit), then every absorbed photon should be balanced by an emitted photon.  Because $\Delta J=+1$ transitions are favored, $R$-branch lines should appear in absorption while $P$-branch lines appear in emission---a phenomenon observed in the $v=1$--0, $\nu_2$ band of water toward Orion~BN/KL \citep{gonzalez_alfonso1998}.  However, the high-mass protostars observed by \citet{boonman2003} show neither this signature nor any clear signs of emission at all.  Of course, astronomical sources are not idealized spherical objects, and a larger amount of material along the line of sight---e.g., the disk/torus structure of AFGL~2136~IRS~1 described in Section \ref{section_source}---would preferentially show absorption.  Additionally, at very high densities collisional de-excitation of vibrationally excited states could compete with spontaneous emission, again favoring absorption being observed.  Our own observations reveal no hint of emission in the $\nu_1$ and $\nu_3$ bands (although we only cover $R$-branch lines), and archival TEXES (Texas Echelon Cross Echelle Spectrograph) observations near 8.5~$\mu$m show the $\nu_2$~12$_{5,8}$--13$_{6,7}$ transition---a $P$-branch line---in absorption.  For all of these reasons, we neglect the possible effects of emission during the analysis of our data.

An alternative probe of density and temperature is CO.  Absorption lines from the $v=1$--0 and $v=2$--1 bands of $^{12}$CO and the $v=1$--0 band of $^{13}$CO near 4.7~$\mu$m were observed toward AFGL~2136~IRS~1 by \citet{mitchell1990co}.  Lines in the $v=1$--0 band of $^{12}$CO are optically thick, so the $^{13}$CO band is used in their main analysis.  \citet{mitchell1990co} find both a cold ($T=17^{+5}_{-3}$~K, $N({\rm ^{13}CO})=(1.2\pm0.5)\times10^{17}$~cm$^{-2}$) and warm ($T=580^{+60}_{-50}$~K, $N({\rm ^{13}CO})=(2.5\pm0.7)\times10^{17}$~cm$^{-2}$) gas component.  The temperature of the warm component is in good agreement with that found by our analysis, and the CO abundance, $N({\rm ^{12}CO})/N({\rm H_2})=(1.8$--$3.2)\times10^{-4}$ assuming $^{12}{\rm CO}/^{13}{\rm CO}=70$ \citep[][and references therein]{sheffer2007}, is indicative of the case where nearly all of the carbon and half of the oxygen are in the form of CO, while the remaining oxygen is in H$_2$O.  Additionally, detection of the $v=2$--1 band of $^{12}$CO in absorption indicates high density ($n({\rm H}_2)\gtrsim10^{10}$~cm$^{-3}$), as an observable population in the first vibrationally excited state would not otherwise be maintained \citep{mitchell1990co}.  This is consistent with the lower limit we have placed on density.

The same set of observations used in our paper also revealed absorption in both the $v=1$--0, $R(0)$ and $R(1)$ transitions of HF at 20.3~km~s$^{-1}$ and 24.5~km~s$^{-1}$, respectively \citep{indriolo2013_HF}.  Given the relative populations in these levels, the velocity match between the $R(1)$ line and the warm H$_2$O, and the characteristic $\sim4$~km~s$^{-1}$ shift between the two lines, we concluded that the $R(0)$ absorption arises in the cold molecular cloud, while the $R(1)$ absorption arises in the same warm gas where we see water.  Significant population in the $J=1$ level requires either radiative pumping or high density, $n({\rm H}_2)\gtrsim10^{9}$~cm$^{-3}$ \citep{neufeld1997}, so the HF results indicate conditions similar to those inferred from H$_2$O and CO.  

Kinematics in the gas surrounding AFGL~2136~IRS~1 are also of interest.  As mentioned before, there is a slight shift between the velocity found for warm $^{13}$CO ($v_{\rm LSR}=26.5$~km~s$^{-1}$) and the systemic velocity ($v_{\rm LSR}=22.8$~km~s$^{-1}$) of the cold cloud surrounding the protostar \citep{kastner1994,vandertak2000yso}.  \citet{kastner1994} suggested that the difference in velocity may indicate that the warm gas is infalling toward the central protostar.  The average velocity found for warm H$_2$O ($v_{\rm LSR}=24.6$~km~s$^{-1}$) is about halfway between the cold cloud and warm $^{13}$CO velocities, a somewhat puzzling result.  Water maser emission at 22.23508~GHz ($6_{1,6}\rightarrow5_{2,3}$) has been observed toward AFGL~2136 in several instances \citep{benson1996,valdettaro2001,menten2004,sunada2007}, with the peak emission between 26.7~km~s$^{-1}$ and 27.5~km~s$^{-1}$.  The interferometric observations of \citet{menten2004} reveal compact maser emission in a region 600$\times$1000~AU around what they dub radio source 4 (RS4), which is coincident with IRS~1 within positional uncertainties.  Conditions necessary to produce this maser emission \citep[$T\sim400$~K, $n({\rm H}_2)\sim10^9$~cm$^{-3}$;][]{elitzur1989} are similar to those that we have inferred, and \citet{menten2004} speculated that the H$_2$O masers may arise in the accretion shock where gas is falling onto the central protostar.  Maser emission, H$_2$O absorption, and CO absorption all paint a consistent picture of warm, dense gas very close to and falling toward, if not onto, the central source.

\subsection{Comparison with Model Results}

As previously stated, the relative abundance of H$_2$O with respect to H$_2$ in the warm gas in protostellar envelopes is expected to be $10^{-4}$.  The reason for this high water abundance is twofold.  First, when the dust temperature is above about 100~K all of the water ice evaporates, drastically increasing the gas-phase H$_2$O abundance.  Second, when the gas temperature is above about 300~K neutral-neutral reactions of H$_2$ with O and OH rapidly drive all of the remaining free oxygen into H$_2$O.  Chemical models that account for changes in the physical structure of protostellar systems with time predict that as time progresses larger portions of the envelope will be at high temperature \citep{ceccarelli1996}, thus increasing the total gas-phase water abundance and the ratio of gas-to-solid H$_2$O.  This suggests the gas-to-solid water ratio as a potential tracer of protostellar evolution.  Indeed, \citet{boonman2003} find that larger H$_2$O abundances and larger gas-to-solid ratios correlate with higher temperatures, the presence of molecules that trace more evolved sources, and evidence of thermal processing in ices.  Our water observations are consistent with the picture of AFGL~2136~IRS~1 as a more evolved protostar, and the increased water column density with respect to that reported by \citet{boonman2003} brings the gas-to-solid ratio into better agreement with what those authors found for AFGL~2591 and AFGL~4176, two other highly evolved sources.

It is also interesting to compare our results to physical models of protostellar envelopes.  \citet{vandertak2000yso} used dust emission and molecular line emission maps to infer density profiles of the form $n(r)=n_{0}(r/r_{0})^{-\alpha}$ for several massive protostars.  In the case of AFGL~2136 they found $r_{0}=35000$~AU, $n_{0}=3.6\times10^4$~cm$^{-3}$, and $\alpha=1.25$, with the model applicable for radii in the range $r_{0}/150\leq r\leq 2r_{0}$.  The density at the inner radius (233~AU) is about $2\times10^{7}$~cm$^{-3}$, much smaller than the lower limit we infer, and a path length of greater than a few hundred AU is necessary to reach a column density comparable to that quoted above for warm H$_2$.   However, they also found unresolved compact emission at 86~GHz indicative of a warm dust shell or circumstellar disk with radius $\lesssim300$~AU.  Such a region seems a favorable place to give rise to the water absorption features that we observe.

\section{FUTURE PROSPECTS}
The large water column density, $N({\rm H_2O})\approx10^{19}$~cm$^{-2}$, and warm temperature, $T\approx500$~K, near AFGL~2136~IRS~1 combine to make absorption lines in the $\nu_1$ and $\nu_3$ bands observable.  To determine the utility of such observations in other sources, we briefly examined predicted line strengths given different values of $T$ and $N({\rm H_2O})$ assuming LTE.  Given the noise level in our spectrum, at 400~K about 75\% of the transitions we detected would still be observable, while at 300~K only about 25\% would remain observable.    Similarly, a reduction in the water column density to $10^{18}$~cm$^{-2}$ drops the number of detecable transitions to about 25\% of those observed.  It is no surprise then that our identical observations of Elias~29---where $N({\rm H_2O})=(7\pm4)\times10^{17}$~cm$^{-2}$; $T=350\pm200$~K \citep{boogert2000}---yield no detections of water lines in absorption.  On the other hand, our initial observations of AFGL~4176 \citep[$N({\rm H_2O})=(1.5\pm0.7)\times10^{18}$~cm$^{-2}$; $T=400\pm250$~K;][]{boonman2003} do show water absorption near 2.5~$\mu$m (to be presented in a future publication).  Clearly, there is a range of temperatures and water column densities over which these observations in the near-IR are currently feasible.  The twelve massive protostars studied by \citet{boonman2003} represent a good sample of sources for initial consideration, with AFGL~2591, NGC~3576, and AFGL~2059 being particularly favorable targets beyond the two already observed.  As telescopes and detectors continue to improve the $\nu_1$ and $\nu_3$ bands of water should become useful tracers of physical conditions in a wide variety of star-forming regions.

\section{SUMMARY} \label{section_sum}
Our observations represent the first extensive study of water absorption in the near-infrared.  Assigning absorption features to 47 separate transitions, we constructed a rotation diagram which is indicative of LTE for levels as high as 4294.5~K above ground.  A statistical equilibrium analysis of all level populations gives best-fit parameters of $T=506\pm25$~K, $n({\rm H_2})>5\times10^9$~cm$^{-3}$, and $N({\rm H_2O})=(1.02\pm0.02)\times10^{19}$~cm$^{-2}$.  These conditions are consistent with those inferred from observations of both CO in absorption and 22~GHz H$_2$O maser emission.  Warm, very dense gas is in close proximity to AFGL~2136~IRS~1 and is likely falling toward, if not onto the massive protostar.

The authors thank the anonymous referee for suggestions to improve the clarity of the paper.  N.I. and D.A.N. are funded by NASA Research Support Agreement No. 1393741 provided through JPL.  M.J.R. is supported by NASA Collaborative Agreement NNX13ai85a.


\begin{thebibliography}{41}
\expandafter\ifx\csname natexlab\endcsname\relax\def\natexlab#1{#1}\fi

\bibitem[{{Benson} \& {Little-Marenin}(1996)}]{benson1996}
{Benson}, P.~J., \& {Little-Marenin}, I.~R. 1996, \apjs, 106, 579

\bibitem[{{Boogert} {et~al.}(2000){Boogert}, {Tielens}, {Ceccarelli},
  {Boonman}, {van Dishoeck}, {Keane}, {Whittet}, \& {de Graauw}}]{boogert2000}
{Boogert}, A.~C.~A., {Tielens}, A.~G.~G.~M., {Ceccarelli}, C., {et~al.} 2000,
  \aap, 360, 683

\bibitem[{{Boonman} \& {van Dishoeck}(2003)}]{boonman2003}
{Boonman}, A.~M.~S., \& {van Dishoeck}, E.~F. 2003, \aap, 403, 1003

\bibitem[{{Brooke} {et~al.}(1999){Brooke}, {Sellgren}, \&
  {Geballe}}]{brooke1999}
{Brooke}, T.~Y., {Sellgren}, K., \& {Geballe}, T.~R. 1999, \apj, 517, 883

\bibitem[{{Carr} {et~al.}(2004){Carr}, {Tokunaga}, \& {Najita}}]{carr2004}
{Carr}, J.~S., {Tokunaga}, A.~T., \& {Najita}, J. 2004, \apj, 603, 213

\bibitem[{{Ceccarelli} {et~al.}(1996){Ceccarelli}, {Hollenbach}, \&
  {Tielens}}]{ceccarelli1996}
{Ceccarelli}, C., {Hollenbach}, D.~J., \& {Tielens}, A.~G.~G.~M. 1996, \apj,
  471, 400

\bibitem[{{Cernicharo} {et~al.}(1990){Cernicharo}, {Thum}, {Hein}, {John},
  {Garcia}, \& {Mattioco}}]{cernicharo1990}
{Cernicharo}, J., {Thum}, C., {Hein}, H., {et~al.} 1990, \aap, 231, L15

\bibitem[{{Cheung} {et~al.}(1969){Cheung}, {Rank}, {Townes}, {Thornton}, \&
  {Welch}}]{cheung1969}
{Cheung}, A.~C., {Rank}, D.~M., {Townes}, C.~H., {Thornton}, D.~D., \& {Welch},
  W.~J. 1969, \nat, 221, 626

\bibitem[{{Daniel} {et~al.}(2011){Daniel}, {Dubernet}, \&
  {Grosjean}}]{daniel2011}
{Daniel}, F., {Dubernet}, M.-L., \& {Grosjean}, A. 2011, \aap, 536, A76

\bibitem[{{de Wit} {et~al.}(2011){de Wit}, {Hoare}, {Oudmaijer},
  {N{\"u}rnberger}, {Wheelwright}, \& {Lumsden}}]{dewit2011}
{de Wit}, W.~J., {Hoare}, M.~G., {Oudmaijer}, R.~D., {et~al.} 2011, \aap, 526,
  L5

\bibitem[{{Doty} {et~al.}(2002){Doty}, {van Dishoeck}, {van der Tak}, \&
  {Boonman}}]{doty2002}
{Doty}, S.~D., {van Dishoeck}, E.~F., {van der Tak}, F.~F.~S., \& {Boonman},
  A.~M.~S. 2002, \aap, 389, 446

\bibitem[{{Elitzur} {et~al.}(1989){Elitzur}, {Hollenbach}, \&
  {McKee}}]{elitzur1989}
{Elitzur}, M., {Hollenbach}, D.~J., \& {McKee}, C.~F. 1989, \apj, 346, 983

\bibitem[{{Gerakines} {et~al.}(1999){Gerakines}, {Whittet}, {Ehrenfreund},
  {Boogert}, {Tielens}, {Schutte}, {Chiar}, {van Dishoeck}, {Prusti},
  {Helmich}, \& {de Graauw}}]{gerakines1999}
{Gerakines}, P.~A., {Whittet}, D.~C.~B., {Ehrenfreund}, P., {et~al.} 1999,
  \apj, 522, 357

\bibitem[{{Gonzalez-Alfonso} {et~al.}(1998){Gonzalez-Alfonso}, {Cernicharo},
  {van Dishoeck}, {Wright}, \& {Heras}}]{gonzalez_alfonso1998}
{Gonzalez-Alfonso}, E., {Cernicharo}, J., {van Dishoeck}, E.~F., {Wright},
  C.~M., \& {Heras}, A. 1998, \apjl, 502, L169

\bibitem[{{Indriolo} {et~al.}(2013){Indriolo}, {Neufeld}, {Seifahrt}, \&
  {Richter}}]{indriolo2013_HF}
{Indriolo}, N., {Neufeld}, D.~A., {Seifahrt}, A., \& {Richter}, M.~J. 2013,
  \apj, 764, 188

\bibitem[{{Kastner} {et~al.}(1992){Kastner}, {Weintraub}, \&
  {Aspin}}]{kastner1992}
{Kastner}, J.~H., {Weintraub}, D.~A., \& {Aspin}, C. 1992, \apj, 389, 357

\bibitem[{{Kastner} {et~al.}(1994){Kastner}, {Weintraub}, {Snell}, {Sandell},
  {Aspin}, {Hughes}, \& {Baas}}]{kastner1994}
{Kastner}, J.~H., {Weintraub}, D.~A., {Snell}, R.~L., {et~al.} 1994, \apj, 425,
  695

\bibitem[{{K\"{a}ufl} {et~al.}(2004){K\"{a}ufl}, {Ballester}, {Biereichel},
  {Delabre}, {Donaldson}, {Dorn}, {Fedrigo}, {Finger}, {Fischer}, {Franza},
  {Gojak}, {Huster}, {Jung}, {Lizon}, {Mehrgan}, {Meyer}, {Moorwood}, {Pirard},
  {Paufique}, {Pozna}, {Siebenmorgen}, {Silber}, {Stegmeier}, \&
  {Wegerer}}]{kaufl2004}
{K\"{a}ufl}, H., {Ballester}, P., {Biereichel}, P., {et~al.} 2004, \procspie,
  5492, 1218

\bibitem[{{Melnick} {et~al.}(1993){Melnick}, {Menten}, {Phillips}, \&
  {Hunter}}]{melnick1993}
{Melnick}, G.~J., {Menten}, K.~M., {Phillips}, T.~G., \& {Hunter}, T. 1993,
  \apjl, 416, L37

\bibitem[{{Menten} {et~al.}(1990{\natexlab{a}}){Menten}, {Melnick}, \&
  {Phillips}}]{menten1990_321GHz}
{Menten}, K.~M., {Melnick}, G.~J., \& {Phillips}, T.~G. 1990{\natexlab{a}},
  \apjl, 350, L41

\bibitem[{{Menten} {et~al.}(1990{\natexlab{b}}){Menten}, {Melnick}, {Phillips},
  \& {Neufeld}}]{menten1990_325GHz}
{Menten}, K.~M., {Melnick}, G.~J., {Phillips}, T.~G., \& {Neufeld}, D.~A.
  1990{\natexlab{b}}, \apjl, 363, L27

\bibitem[{{Menten} \& {van der Tak}(2004)}]{menten2004}
{Menten}, K.~M., \& {van der Tak}, F.~F.~S. 2004, \aap, 414, 289

\bibitem[{{Minchin} {et~al.}(1991){Minchin}, {Hough}, {Burton}, \&
  {Yamashita}}]{minchin1991}
{Minchin}, N.~R., {Hough}, J.~H., {Burton}, M.~G., \& {Yamashita}, T. 1991,
  \mnras, 251, 522

\bibitem[{{Mitchell} {et~al.}(1990){Mitchell}, {Maillard}, {Allen}, {Beer}, \&
  {Belcourt}}]{mitchell1990co}
{Mitchell}, G.~F., {Maillard}, J.-P., {Allen}, M., {Beer}, R., \& {Belcourt},
  K. 1990, \apjs, 363, 554

\bibitem[{{Murakawa} {et~al.}(2008){Murakawa}, {Preibisch}, {Kraus}, \&
  {Weigelt}}]{murakawa2008}
{Murakawa}, K., {Preibisch}, T., {Kraus}, S., \& {Weigelt}, G. 2008, \aap, 490,
  673

\bibitem[{{Najita} {et~al.}(2000){Najita}, {Edwards}, {Basri}, \&
  {Carr}}]{najita2000}
{Najita}, J.~R., {Edwards}, S., {Basri}, G., \& {Carr}, J. 2000, Protostars and
  Planets IV, 457

\bibitem[{{Neufeld}(2010)}]{neufeld2010_nn}
{Neufeld}, D.~A. 2010, \apj, 708, 635

\bibitem[{{Neufeld} {et~al.}(1997){Neufeld}, {Zmuidzinas}, {Schilke}, \&
  {Phillips}}]{neufeld1997}
{Neufeld}, D.~A., {Zmuidzinas}, J., {Schilke}, P., \& {Phillips}, T.~G. 1997,
  \apjl, 488, L141

\bibitem[{{Pontoppidan} {et~al.}(2010{\natexlab{a}}){Pontoppidan}, {Salyk},
  {Blake}, \& {K{\"a}ufl}}]{pontoppidan2010}
{Pontoppidan}, K.~M., {Salyk}, C., {Blake}, G.~A., \& {K{\"a}ufl}, H.~U.
  2010{\natexlab{a}}, \apjl, 722, L173

\bibitem[{{Pontoppidan} {et~al.}(2010{\natexlab{b}}){Pontoppidan}, {Salyk},
  {Blake}, {Meijerink}, {Carr}, \& {Najita}}]{pontoppidan2010spitzer}
{Pontoppidan}, K.~M., {Salyk}, C., {Blake}, G.~A., {et~al.} 2010{\natexlab{b}},
  \apj, 720, 887

\bibitem[{{Salyk} {et~al.}(2008){Salyk}, {Pontoppidan}, {Blake}, {Lahuis}, {van
  Dishoeck}, \& {Evans}}]{salyk2008}
{Salyk}, C., {Pontoppidan}, K.~M., {Blake}, G.~A., {et~al.} 2008, \apjl, 676,
  L49

\bibitem[{{Seifahrt} {et~al.}(2010){Seifahrt}, {K{\"a}ufl}, {Z{\"a}ngl},
  {Bean}, {Richter}, \& {Siebenmorgen}}]{seifahrt2010}
{Seifahrt}, A., {K{\"a}ufl}, H.~U., {Z{\"a}ngl}, G., {et~al.} 2010, \aap, 524,
  A11

\bibitem[{{Sheffer} {et~al.}(2007){Sheffer}, {Rogers}, {Federman}, {Lambert},
  \& {Gredel}}]{sheffer2007}
{Sheffer}, Y., {Rogers}, M., {Federman}, S.~R., {Lambert}, D.~L., \& {Gredel},
  R. 2007, \apj, 667, 1002

\bibitem[{{Sonnentrucker} {et~al.}(2010){Sonnentrucker}, {Neufeld}, {Phillips},
  {Gerin}, {Lis}, {de Luca}, {Goicoechea}, {Black}, {Bell}, {Boulanger},
  {Cernicharo}, {Coutens}, {Dartois}, {Ka{\'z}mierczak}, {Encrenaz},
  {Falgarone}, {Geballe}, {Giesen}, {Godard}, {Goldsmith}, {Gry}, {Gupta},
  {Hennebelle}, {Herbst}, {Hily-Blant}, {Joblin}, {Ko{\l}os}, {Kre{\l}owski},
  {Mart{\'{\i}}n-Pintado}, {Menten}, {Monje}, {Mookerjea}, {Pearson},
  {Perault}, {Persson}, {Plume}, {Salez}, {Schlemmer}, {Schmidt}, {Stutzki},
  {Teyssier}, {Vastel}, {Yu}, {Caux}, {G{\"u}sten}, {Hatch}, {Klein}, {Mehdi},
  {Morris}, \& {Ward}}]{sonnentrucker2010}
{Sonnentrucker}, P., {Neufeld}, D.~A., {Phillips}, T.~G., {et~al.} 2010, \aap,
  521, L12

\bibitem[{{Sunada} {et~al.}(2007){Sunada}, {Nakazato}, {Ikeda}, {Hongo},
  {Kitamura}, \& {Yang}}]{sunada2007}
{Sunada}, K., {Nakazato}, T., {Ikeda}, N., {et~al.} 2007, \pasj, 59, 1185

\bibitem[{{Valdettaro} {et~al.}(2001){Valdettaro}, {Palla}, {Brand},
  {Cesaroni}, {Comoretto}, {Di Franco}, {Felli}, {Natale}, {Palagi}, {Panella},
  \& {Tofani}}]{valdettaro2001}
{Valdettaro}, R., {Palla}, F., {Brand}, J., {et~al.} 2001, \aap, 368, 845

\bibitem[{{van der Tak} {et~al.}(2000){van der Tak}, {van Dishoeck}, {Evans},
  \& {Blake}}]{vandertak2000yso}
{van der Tak}, F.~F.~S., {van Dishoeck}, E.~F., {Evans}, II, N.~J., \& {Blake},
  G.~A. 2000, \apj, 537, 283

\bibitem[{{van Dishoeck} \& {Helmich}(1996)}]{vandishoeck1996}
{van Dishoeck}, E.~F., \& {Helmich}, F.~P. 1996, \aap, 315, L177

\bibitem[{{van Dishoeck} {et~al.}(1996){van Dishoeck}, {Helmich}, {de Graauw},
  {Black}, {Boogert}, {Ehrenfreund}, {Gerakines}, {Lacy}, {Millar}, {Schutte},
  {Tielens}, {Whittet}, {Boxhoorn}, {Kester}, {Leech}, {Roelfsema}, {Salama},
  \& {Vandenbussche}}]{vandishoeck1996CO2}
{van Dishoeck}, E.~F., {Helmich}, F.~P., {de Graauw}, T., {et~al.} 1996, \aap,
  315, L349

\bibitem[{{van Dishoeck} {et~al.}(2011){van Dishoeck}, {Kristensen}, {Benz},
  {Bergin}, {Caselli}, {Cernicharo}, {Herpin}, {Hogerheijde}, {Johnstone},
  {Liseau}, {Nisini}, {Shipman}, {Tafalla}, {van der Tak}, {Wyrowski},
  {Aikawa}, {Bachiller}, {Baudry}, {Benedettini}, {Bjerkeli}, {Blake},
  {Bontemps}, {Braine}, {Brinch}, {Bruderer}, {Chavarr{\'{\i}}a}, {Codella},
  {Daniel}, {de Graauw}, {Deul}, {di Giorgio}, {Dominik}, {Doty}, {Dubernet},
  {Encrenaz}, {Feuchtgruber}, {Fich}, {Frieswijk}, {Fuente}, {Giannini},
  {Goicoechea}, {Helmich}, {Herczeg}, {Jacq}, {J{\o}rgensen}, {Karska},
  {Kaufman}, {Keto}, {Larsson}, {Lefloch}, {Lis}, {Marseille}, {McCoey},
  {Melnick}, {Neufeld}, {Olberg}, {Pagani}, {Pani{\'c}}, {Parise}, {Pearson},
  {Plume}, {Risacher}, {Salter}, {Santiago-Garc{\'{\i}}a}, {Saraceno},
  {St{\"a}uber}, {van Kempen}, {Visser}, {Viti}, {Walmsley}, {Wampfler}, \&
  {Y{\i}ld{\i}z}}]{vandishoeck2011}
{van Dishoeck}, E.~F., {Kristensen}, L.~E., {Benz}, A.~O., {et~al.} 2011,
  \pasp, 123, 138

\bibitem[{{Willner} {et~al.}(1982){Willner}, {Gillett}, {Herter}, {Jones},
  {Krassner}, {Merrill}, {Pipher}, {Puetter}, {Rudy}, {Russell}, \&
  {Soifer}}]{willner1982}
{Willner}, S.~P., {Gillett}, F.~C., {Herter}, T.~L., {et~al.} 1982, \apj, 253,
  174

\end{thebibliography}


\clearpage


\clearpage


\clearpage


\clearpage


\clearpage

\begin{deluxetable}{clrccccccclc}
\tablecolumns{12}
\tablewidth{8.5in}
\rotate
\tabletypesize{\scriptsize}
\tablecaption{Absorption Line Parameters \label{tbl_absorption}}
\tablehead{\colhead{Wavelength} & \colhead{} & \colhead{$E/k$} & \colhead{$W_{\lambda}$-Pred.} & \colhead{$W_{\lambda}$-Obs.} & \colhead{$\sigma(W_{\lambda})$} & \colhead{$v_{\rm LSR}$}  & \colhead{FWHM} & \colhead{$N(J_{K_{a},K_{c}})$} &  \colhead{$\sigma(N)$} & \colhead{Figure \ref{fig_spectra}} & \colhead{} \\
\colhead{($\mu$m)} & \colhead{Transition}  & \colhead{(K)} & \colhead{($10^{-6}$~$\mu$m)} & \colhead{($10^{-6}$~$\mu$m)} & \colhead{($10^{-6}$~$\mu$m)}    & \colhead{(km~s$^{-1}$)} & \colhead{(km~s$^{-1}$)} & \colhead{($10^{15}$~cm$^{-2}$)} & \colhead{($10^{15}$~cm$^{-2}$)} & \colhead{Panel} & \colhead{Comments}
}
\startdata
2.4686156 & $\nu_{1}$~12$_{6,7}$--11$_{5,6}$     & 2879.4 & 1.95 & 1.89 & 0.20 & 25.6 & 14.4 & 5.52 & 0.57 & A1  & ... \\
2.4689113 & $\nu_{3}$~13$_{4,9}$--12$_{4,8}$     & 3177.1 & 1.57 & 1.79 & 0.12 & 24.5 & 13.0 & 1.30 & 0.09 & A2  & ... \\
2.4732020 & $\nu_{1}$~11$_{6,5}$--10$_{5,6}$     & 2475.7 & 1.73 & 2.16 & 0.17 & 25.0 & 14.4 & 14.6 & 1.13 & A3  & ... \\
2.4738368 & $\nu_{3}$~14$_{6,8}$--13$_{6,7}$     & 3970.4 & 0.65 & 0.84 & 0.15 & 24.2 & 11.1 & 0.96 & 0.17 & A4  & ... \\
2.4740698 & $\nu_{3}$~18$_{1,18}$--17$_{1,17}$ & 4294.5 & 0.23 & ...     & ...     & ...     & ....    & 0.19\tablenotemark{b} & 0.02\tablenotemark{b} & A5  & blend 1 \\
2.4740698 & $\nu_{3}$~18$_{0,18}$--17$_{0,17}$ & 4294.5 & 0.70 & 1.04\tablenotemark{a} & 0.12\tablenotemark{a} & 24.3 & 10.8 & 0.57\tablenotemark{b} & 0.06\tablenotemark{b} & A5*  & blend 1 \\
2.4748404 & $\nu_{3}$~15$_{4,12}$--14$_{4,11}$ & 3955.5 & 1.04 & 1.25 & 0.14 & 24.2 & 11.7 & 1.01 & 0.11 & B1  & ... \\
2.4752485 & $\nu_{1}$~11$_{5,6}$--10$_{4,7}$     & 2277.8 & 2.00 & 1.05 & 0.11 & 23.7 & 10.0 & 9.10 & 0.95 & B2  & ... \\
2.4757180 & $\nu_{1}$~10$_{7,4}$--9$_{6,3}$       & 2349.9 & 1.18 & 2.11 & 0.34 & 24.8 & 17.6 & 24.3 & 3.87 & B3  & HF $R(1)$ \\ 
2.4774810 & $\nu_{3}$~14$_{3,11}$--13$_{3,10}$ & 3478.3 & 2.70 & 3.14 & 0.13 & 24.7 & 13.9 & 2.35 & 0.09 & B4  & ... \\ 
2.4785984 & $\nu_{3}$~12$_{4,8}$--11$_{4,7}$     & 2735.4 & 10.8 & 8.75\tablenotemark{c} & 0.46\tablenotemark{c} & ...     & ...     & 6.15 & 0.32 & B5  & ... \\ 
2.4833761 & $\nu_{3}$~13$_{3,10}$--12$_{3,9}$   & 3033.4 & 2.11 & 1.36 & 0.14 & 24.5 & 11.5 & 0.98 & 0.10 & C1  & ... \\ 
2.4861802 & $\nu_{1}$~10$_{6,4}$--9$_{5,5}$       & 2124.6 & 0.87 & 1.11 & 0.17 & 23.0 & 17.1 & 9.10 & 1.39 & C2  & ... \\ 
2.4868917 & $\nu_{1}$~7$_{4,4}$--6$_{1,5}$         &   782.0 & 1.17 & 1.78 & 0.22 & 24.8 & 14.9 & 108  & 13.2 & C3  & ... \\ 
2.4887954 & $\nu_{2}$~15$_{3,12}$--14$_{2,13}$ & 3353.2 & 1.31 & 2.10 & 0.13 & 24.9 & 12.2 & 4.47 & 0.28 & C4  & ... \\ 
2.4893977 & $\nu_{3}$~12$_{3,9}$--11$_{3,8}$     & 2611.8 & 14.3 & 12.7 & 0.49 & 23.1 & 14.5 & 8.66 & 0.34 & C5  & ... \\ 
2.4895351 & $\nu_{3}$~14$_{2,12}$--13$_{2,11}$ & 3236.5 & 4.64 & ...     & ...     & ...     & ...     & $<10.7$ & ... & D1  & blend 2 \\ 
2.4895765 & $\nu_{3}$~16$_{0,16}$--15$_{0,15}$ & 3397.0 & 4.13 & 15.2\tablenotemark{a} & 0.55\tablenotemark{a} & ...     & ...     & $<10.0$ & ... & D1*  & blend 2 \\  
2.4895766 & $\nu_{3}$~16$_{1,16}$--15$_{1,15}$ & 3397.0 & 1.38 & ...     & ...     & ...     & ...     & $<10.0$ & ... & D1  & blend 2 \\  
2.4895769 & $\nu_{3}$~13$_{5,9}$--12$_{5,8}$     & 3277.5 & 3.09 & ...     & ...     & ...     & ...     & $<13.7$ & ... & D1  & blend 2 \\  
2.4901296 & $\nu_{3}$~14$_{3,12}$--13$_{3,11}$ & 3238.2 & 1.54 & 0.95 & 0.14 & 23.6 & 10.5 & 0.67 & 0.10 & D2  & ... \\  
2.4904204 & $\nu_{3}$~11$_{4,7}$--10$_{4,6}$     & 2328.4 & 7.19 & 6.56 & 0.39 & 24.7 & 14.9 & 4.75 & 0.28 & D3  & ... \\  
2.4908999 & $\nu_{3}$~15$_{2,14}$--14$_{2,13}$ & 3353.2 & 2.87 & 2.70 & 2.00 & 23.4 & 15.2 & 2.61 & 1.93 & D4  & blend 3 \\  
2.4909997 & $\nu_{3}$~12$_{5,7}$--11$_{5,6}$     & 2879.4 & 5.41 & 5.85 & 1.98 & 23.9 & 15.2 & 6.16 & 2.08 & D4*  & blend 3 \\  
2.4912320 & $\nu_{3}$~13$_{4,10}$--12$_{4,9}$   & 3060.9 & 5.70 & 5.72 & 0.16 & 24.7 & 14.8 & 4.32 & 0.12 & D5  & ... \\  
2.4977585 & $\nu_{3}$~15$_{0,15}$--14$_{0,14}$ & 2986.8 & 3.08 & ...     & ...     & ...     & ...     & 1.53\tablenotemark{b} & 0.07\tablenotemark{b} & E1  & blend 4 \\  
2.4977590 & $\nu_{3}$~15$_{1,15}$--14$_{1,14}$ & 2986.8 & 9.22 & 9.74\tablenotemark{a} & 0.43\tablenotemark{a} & 25.0 & 14.4 & 4.59\tablenotemark{b} & 0.20\tablenotemark{b} & E1*  & blend 4 \\  
2.4981464 & $\nu_{3}$~13$_{3,11}$--12$_{3,10}$ & 2826.9 & 10.2 & 25.4\tablenotemark{a} & 1.24\tablenotemark{a} & ...     & ...     & $<17.2$ & ... & E2*  & blend 5 \\  
2.4982130 & $\nu_{3}$~14$_{1,13}$--13$_{1,12}$ & 2941.8 & 9.06 & ...     & ...     & ...     & ...     &  $<16.5$ & ... & E2  & blend 5 \\  
2.4982710 & $\nu_{3}$~14$_{2,13}$--13$_{2,12}$ & 2941.9 & 3.02 & ...     & ...     & ...     & ...     &  $<16.5$ & ... & E2  & blend 5 \\  
2.4991990 & $\nu_{1}$~9$_{6,3}$--8$_{5,4}$         & 1808.0 & 3.18 & 3.00 & 0.20 & 23.7 & 13.4 & 33.8 & 2.30 & E3  & HF $R(0)$ \\  
2.4997063 & $\nu_{1}$~9$_{6,4}$--8$_{5,3}$         & 1809.1 & 1.12 & 1.82 & 0.16 & 23.9 & 15.9 & 19.3 & 1.74 & E4  & ... \\  
2.5006937 & $\nu_{3}$~12$_{5,8}$--11$_{5,7}$     & 2860.4 & 2.22 & ...     & ...     & ...     & ...     & $<5.05$ & ... & E5  & blend 6 \\  
2.5007295 & $\nu_{3}$~12$_{4,9}$--11$_{4,8}$     & 2654.8 & 4.01 & 5.70\tablenotemark{a} & 0.19\tablenotemark{a} & ...     & ...     & $<4.22$ & ... & E5*  & blend 6 \\  
2.5018269 & $\nu_{1}$~8$_{7,1}$--7$_{6,2}$         & 1751.8 & 0.35 & ...     & ...     & ...     & ...     & 18.5\tablenotemark{b} & 10.6\tablenotemark{b} & F1  & blend 7 \\  
2.5018296 & $\nu_{1}$~8$_{7,2}$--7$_{6,1}$         & 1751.8 & 1.06 & 2.21\tablenotemark{a} & 1.27\tablenotemark{a} & 22.7 & 19.0 & 55.5\tablenotemark{b} & 31.9\tablenotemark{b} & F1*  & blend 7 \\  
2.5045908 & $\nu_{3}$~12$_{2,10}$--11$_{2,9}$   & 2435.3 & 21.8 & 30.8\tablenotemark{c} & 0.68\tablenotemark{c} & ...     & ...     & 16.8 & 0.38 & F2  & ... \\  
2.5066038 & $\nu_{3}$~12$_{3,10}$--11$_{3,9}$   & 2441.6 & 6.98 & ...     & ...     & ...     & ...     & $<15.8$ & ... & F3  & blend 8 \\  
2.5066091 & $\nu_{3}$~13$_{2,12}$--12$_{2,11}$ & 2556.4 & 19.0 & 24.3\tablenotemark{a} & 0.37\tablenotemark{a} & ...     & ...     & $<15.1$ & ... & F3*  & blend 8 \\  
2.5112606 & $\nu_{3}$~11$_{4,8}$--10$_{4,7}$     & 2277.8 & 23.6 & 19.0 & 0.56 & 25.7 & 16.2 & 14.0 & 0.41 & F4  & ... \\  
2.5120359 & $\nu_{3}$~11$_{2,9}$--10$_{2,8}$     & 2071.3 & 14.0 & 10.8 & 0.50 & 25.5 & 14.2 & 6.71 & 0.31 & F5  & ... \\  
2.5126941 & $\nu_{1}$~8$_{6,3}$--7$_{5,2}$         & 1526.6 & 3.10 & ...     & ...     & ...     & ...     & $<273$ & ... & G1  & blend 9 \\  
2.5127076 & $\nu_{3}$~11$_{5,7}$--10$_{5,6}$     & 2475.7 & 13.4 & 15.2\tablenotemark{a} & 0.44\tablenotemark{a} & ...     & ...     & $<13.2$ & ... & G1*  & blend 9 \\  
2.5133118 & $\nu_{1}$~7$_{2,5}$--6$_{1,6}$         &   644.2 & 0.61 & 1.07 & 0.17 & 27.3 &  9.6  & 497  & 78.6 & G2  & ... \\  
2.5184602 & $\nu_{3}$~11$_{6,6}$--10$_{6,5}$     & 2700.8 & 7.38 & 5.61 & 0.30 & 25.3 & 14.6 & 5.64 & 0.30 & G3  & ... \\  
2.5204395 & $\nu_{1}$~6$_{3,4}$--5$_{0,5}$         &   468.6 & 0.70 & 0.65 & 0.21 & 27.8 &  8.6  & 315  & 102  & G4  & ... \\  
2.5209404 & $\nu_{3}$~12$_{8,4}$--11$_{8,3}$     & 3633.2 & 0.80 & 1.29 & 0.25 & 25.0 & 11.6 & 2.04 & 0.40 & G5  & ...  \\
\enddata 
\tablecomments{Columns 1, 2, and 3 give the transition wavelength, label, and lower state energy.  Column 4 is the predicted equivalent width, $W_\lambda$-Pred, assuming a total water column density of $10^{19}$~cm$^{-2}$ in LTE at 500~K.  Columns 5 and 6 are the observed equivalent width, $W_\lambda$-Obs, and its $1\sigma$ statistical uncertainty returned by the fitting procedure.  Systematic uncertainties---due to, e.g., working around poorly removed atmospheric features and setting the continuum level---may be somewhat larger.  Columns 7 and 8 are the line-center velocity and velocity full-width at half-maximum (including instrumental broadening effects) found by a Gaussian fit to the absorption feature.  In cases where blended lines were fit with a single Gaussian component or where features with shoulders were fit with two Gaussian components, velocity information is not reported.  Columns 9 and 10 are the column density in the lower state, $N$, and its $1\sigma$ uncertainty.  Upper limits on column densities are calculated assuming the entire equivalent width of a blended feature is due only to each contributing line in turn.  Column 11 gives the panel in Figure \ref{fig_spectra} in which the absorption feature is shown.  For blended lines, an asterisk next to the panel label denotes the transition wavelength used to set zero velocity on the bottom axis.  Note that beyond the 47 detected transitions that are tabulated above we predict an additional 42 transitions covered by our observations that should be above the detection limit of $W_{\lambda}>6\times10^{-7}$~$\mu$m assuming LTE at 500~K and $N({\rm H_{2}O})=10^{19}$~cm$^{-2}$.  All of these transitions are either obscured by strong telluric lines (39) or affected by bad pixels on the detector (3).  Non-detections are consistent with predicted equivalent widths, and in no case is a transition that we expect to see ``missing''.}
\tablenotetext{a}{Total equivalent width of absorption feature containing multiple blended lines.}
\tablenotetext{b}{Column densities and uncertainties were calculated using the equivalent width of the blended feature and assuming an ortho-to-para ratio of 3:1.}
\tablenotetext{c}{Sum of 2 components used to fit absorption features with shoulders.}
\end{deluxetable}
\normalsize

\end{document}